\documentstyle[10pt,emulateapj]{article}

\begin{document}

\title{The Collapse of Neutron Stars in High-Mass Binaries as
the Energy Source for the Gamma-ray Bursts}

\author{Bo Qin\altaffilmark{1,2}, 
	Xiang-Ping Wu\altaffilmark{1}, 
        Ming-Chung Chu\altaffilmark{2}, 
        Li-Zhi Fang\altaffilmark{3} and
	Jing-Yao Hu\altaffilmark{1} }

\altaffiltext{1}{Beijing Astronomical Observatory, Chinese Academy 
                 of Sciences, Beijing 100080, China}
\altaffiltext{2}{Department of Physics, the Chinese University of Hong
Kong, Shatin, N.T., Hong Kong, China} 
\altaffiltext{3}{Department of Physics, University of Arizona, Tucson, 
                 AZ 85721}

\begin{abstract}
The energy source has remained to be  the great
mystery in understanding of the gamma-ray bursts (GRBs) if
the events are placed at cosmological distances as indicated by 
a number of recent observations. The currently popular models  
include (1)the merger of two neutron stars or a neutron
star and a black hole binary and (2)the hypernova scenario of the
collapse of a massive member in a close binary.  Since a neutron star will
inevitably collapse into a black hole if its mass exceeds 
the limit $M_{max}\approx3M_{\odot}$, 
releasing a total binding gravitational energy of $\sim10^{54}$ erg, 
we explore semi-empirically the possibility of attributing 
the energy source of GRB to the accretion-induced collapse 
of a neutron star (AICNS) in a massive X-ray binary system consisting of
a neutron star and a type O/B companion. This happens because
a significant mass flow of $\sim10^{-3}$--$10^{-4}M_{\odot}$ yr$^{-1}$ 
may be transferred onto the neutron star  through the Roche-lobe overflow
and primarily during the spiral-in phase when it plunges into the
envelope of the companion, which may eventually lead to the AICNS 
before the neutron star merges with the core of the companion.
In this scenario, a ``dirty'' fireball with a moderate amount of beaming 
is naturally expected because of the nonuniformity of the stellar matter
surrounding the explosion inside the companion, 
and a small fraction ($\sim0.1\%$) of the energy is sufficient to 
create the observed GRBs. In addition, the bulk of the ejecting matter
of the companion star with a relatively slow expansion rate 
may act as the afterglow. Assuming a non-evolutionary model for galaxies, 
we estimate that the birthrate of the AICNS events is about 2 per day
within a volume to redshift $z=1$ for an $\Omega_0=1$ universe, 
consistent with the reported GRB rate.  It appears that the AICNS
scenario, as a result of stellar evolution,  
may provide a natural explanation for the origin of  GRBs and therefore
deserves to be further investigated in the theoretical study of GRBs. 
\end{abstract}


\section{Introduction }

The recent detections of the transient counterparts of GRB
970228 and 970508 at optical/radio/X-ray wavelengths (e.g. van Paradijs
et al. 1997; Sahu et al. 1997; Costa et al. 1997; Guarnieri et al. 1997)
and the identification of absorption line associated with the optical 
afterglow of GRB 970508 (Metzger et al. 1997), 
which places the source of the burst at cosmological distance,  
have stimulated a number of authors to re-consider
the theoretical models of the origin of GRBs. The currently popular
model for GRBs is the so-called impulsive ``fireball'' -- a relativistic 
expanding blastwave with a bulk Lorentz factor of a few hundreds, 
releasing energy of the order of $10^{51}$ erg 
(e.g. Cavallo \& Rees 1978; Rees \& M\'esz\'aros, 1992; 
Vietri 1997a,b; Waxman 1997). Two proposed mechanisms remain 
attractive for the energy sources of the fireballs: 
(1) mergers of double neutron star or neutron star and black hole
binaries due to the energy loss of gravitational radiation
(Blinnikov et al. 1984; Eichler et al. 1989; 
Paczy\'nski 1986, 1991, 1992; Rees \& M\'esz\'aros, 1992;
Narayan et al. 1992; Lipunov et al. 1995; Vietri 1997a,b; 
Sahu et al. 1997; Waxman 1997) and (2) 
the ``hypernovae'' -- the gravitational
collapse of a massive star which can be either a ``failed type Ib
supernova'' (Woosley 1993) or a massive member in a close binary system 
(Paczy\'nski 1997). In the latter case, an efficient magnetic energy
transport is required to explode the stellar envelope, releasing an energy 
of $\sim10^{54}$ erg. Although the explosions result from 
different physical processes, both models lead to
the formation of massive neutron stars or black holes.

It is commonly believed that an isolated neutron star is relatively
stable, having a mass less than the limit
$M_{max}\approx3M_{\odot}$. What would happen if there
is a stable mass flow onto a neutron star from, for example, 
its companion in a binary system? When the mass of a neutron star grows
up to $M_{max}$ because of the net mass deposition,
does the neutron star collapse into a black hole? The extreme 
circumstance can be the merger of a binary of double neutron stars with
a total mass of $>M_{max}$, while the most likely case can appear 
in a high-mass X-ray binary where the mass transferred by  
the massive companion to the neutron star may  eventually 
lead the gravitating mass of the neutron star to exceed the maximum 
value $M_{max}$. Such a scenario of accretion-induced collapse of
neutron stars  should be a natural extrapolation of 
stellar evolution,  which is quite similar to the accretion-induced 
collapse of white dwarfs into neutron stars 
(Canal et al. 1990 and references therein). 
The latter was once suggested as a possible source of GRBs
(Goodman, Dar, \& Nussinov 1987; Dar et al. 1992).
If a neutron star collapses into a black hole, the disrupted neutron
star will release catastrophically its gravitational binding energy 
in a very short time, leading to an explosion with energy of
$\sim10^{54}$ erg, namely, the ``fireball''. Are the AICNS in high-mass 
binaries able to act as the energy sources for the GRBs ?

The major difference between the hypernova model and the AICNS model
is the formation of  black holes through the collapse of  
neutron stars. According to the hypernova mechanism, the hot and rapidly
spinning neutron star cools off by neutrino emission to become a 
stellar black hole (Woosley 1993; Paczy\'nski 1997), whereas in
the AICNS scenario the black hole forms because the mass of 
neutron star increases to the point of dynamical instability
($M_{max}$). The similarity between the merging model and the AICNS 
model is that the collapse of a neutron star is caused by the 
external infalling matter.  In a sense, the AICNS as the
origin of GRB is somewhat a combination of the merging model 
and the hypernova model, but seems to be more natural according to
our knowledge of stellar evolution. At least,
the AICNS is a possible scenario that shouldn't be overlooked
in the study of the energy source for GRBs. 
Since the emission mechanism of the GRBs
remains the same as that in the fireball model, we will discuss the GRB
rate and other properties of the AICNS model in this letter.  
As in other proposed models for GRBs, only a semi-empirical approach 
is applicable to such an analysis at the present stage.

\section{The AICNS model}

We confine ourselves to high-mass X-ray binaries consisting of a
neutron stars and a type O or B star companion in order to 
ensure that the material transferred by the companion to 
the neutron star enables the neutron star to reach the point of 
dynamical instability ($M_{max}$). Although a neutron star can
capture a significant fraction of the stellar wind from its companion
with a mass transfer rate 
$\dot{M}=10^{-6}$--$10^{-10}M_{\odot}$ yr$^{-1}$
on the stellar evolution time scale, 
giving rise to the observed X-ray luminosity,  
such a mass accretion rate is insufficient to disrupt the neutron
star within the lifetime of the companion ($\sim10^{6}$ yr). 
A dramatic mass transfer occurs once the companion expands into 
a giant and thereby reaches its Roche lobe.  In particular,   
the mass of the neutron star $M_N$ may rapidly grow up to $M_{max}$
while it 
plunges into the envelope of its companion. Eventually, the AICNS 
may take place before or after the neutron star is engulfed 
by its companion, depending on when the condition $M_N>M_{max}$
is reached. Yet, the AICNS will not take place if the accreted mass of  
a neutron star during the spiral-in to the center of its companion
does not meet $M_N>M_{max}$, which then results in the formation of a 
massive neutron star with $1.4M_{\odot}<M<M_{max}$
instead of a black hole (Verbunt 1993 and references therein).

The kinetic energy of the collapsing matter of a neutron star into a 
black hole can be estimated roughly by 
$E\approx (M_{max}c^2/2)(1-R_{g}/R_N)$,
where $R_N$ is the radius of a neutron star with $M=M_{max}$ and
$R_{g}\equiv 2GM_{max}/c^2$ is the corresponding Schwarzschild radius. 
Taking $M_{max}=2.7M_{\odot}$ and $R_N=13.5$ km (Kippenhahn \& Weigert
1990), we obtain $E\approx1\times10^{54}$ erg. This naturally provides the
energy needed for the explosion and hence for the generation of 
the fireball (see, for example, Paczy\'nski 1997).
If a small fraction of this energy, say $\sim10^{51}$ erg, is 
converted into gamma-ray emission, 
the AICNS scenario will be in good agreement 
with the observed GRB intensity distribution, which shows a standard 
candle peak luminosity of $\sim10^{51}$ erg/s.

It is estimated that there are $N_x\sim50$ massive X-ray binaries in the
Galaxy (Meurs \& van den Heuvel 1989; Bhattacharya \& van den Heuvel
1991) and a significant fraction of 
them are the neutron star-O/B star systems. 
Because the AICNS event rate is closely relevant to the lifetime of the 
O/B stars, which turns out to be $t\approx5\times10^6$ yr for a typical 
$25M_{\odot}$ companion, we can then estimate the GRB birthrate roughly
by $\dot{N}_{Galaxy}\sim N_x/t\approx1\times10^{-5}L_G^{-1}$ yr$^{-1}$,
where $L_G\approx3.6\times10^{10}L_{\odot}$ is the bolometric
luminosity of the Galaxy. This is comparable to the birthrates of 
double neutron star binaries and black hole-neutron star binaries in
the Galaxy (Narayan et al. 1991). Assuming a non-evolutionary scenario
for galaxies and adopting the Schechter luminosity
function with $\phi^*=1.56\times10^{-2}h^3$ Mpc$^{-3}$, $\alpha=-1.1$
and $L^*=7.1\times10^{9}h^{-2}L_{\odot}$ (Fukugita \& Turner 1991 
and references therein), we find the total luminosity of the universe 
within redshift $z=1$ for a cosmological model of $\Omega_0=1$ to be
$L_{total}=2.5\times10^{18}h^{-2}L_{\odot}$, where $h$ is the 
Hubble constant in unit of 100 km/s/Mpc. Finally, the expected GRB event
rate within a volume to $z=1$ is 
$\dot{N}_{GRB}=\dot{N}_{Galaxy} L_{total}\sim700$ yr$^{-1}$, 
i.e., about two events per day.  Such a rate, though rather uncertain,   
is consistent with that recorded by the current detectors (e.g., Meegan
et al. 1996).

\section{Further remarks}

At least $\sim M_{\odot}$ mass of its companion star 
should be accreted onto the neutron star during the spiral-in phase
in order to trigger the AICNS. This includes the final 
merger of the neutron star with the core of the massive companion star 
if the AICNS does not take place before it reaches the core. 
It appears that an average net mass accretion rate 
of $10^{-3}$--$10^{-4}M_{\odot}$ yr$^{-1}$ is required for the neutron
star provided that the spiral-in takes about $10^{3}$--$10^{4}$ years 
(Meurs \& van den Heuvel 1989). 
A simple computation based on the powder approximation (Lipunov 1992)
shows that such a mass accretion rate is indeed possible 
for a neutron star with mass $M\sim M_{\odot}$ 
moving with a velocity $v$ through the medium of mass density $\rho$:
$\dot{M}\sim\pi(2GM_{\odot})^2\rho/v^3
\sim 5\times10^{-3}(v/10^4\;{\rm km/s})^{-3}(\rho/10^{-3}\rho_{\odot})$,
where the average medium density is estimated using a giant star with
mass $30M_{\odot}$ and radius $30R_{\odot}$, the velocity $v$ 
corresponds approximately to the orbital velocity of the neutron star 
during the spiral-in, and 
$\rho_{\odot}=1.4$ g cm$^{-3}$ is the mean mass density of the Sun.  
Apparently, our required mass accretion rate exceeds
the Eddington limit of $1\times 10^{-8} M_{\odot}$yr$^{-1}$
for a neutron star (Taam, Bodenheimer, \& Ostriker 1978; Delgado 1980).
This can only be overcome when neutrino emission is taken into account.
Indeed,  the rapid neutrino cooling may allow 
mass accretion to reach a very high rate onto the neutron 
star. Chevalier (1989) and Houck \& Chevalier (1991) have found that
for steady spherical models the neutron star may have an accretion rate
of $\dot{M} \geq 10^{-3} M_{\odot}$yr$^{-1}$. Chevalier (1993) showed
that a strong accretion is possible for neutron star spiral-in through the 
massive envelope because of neutrino emission, which can bring 
an initial $1.4 M_{\odot}$ neutron star to a massive object of 
$>2 M_{\odot}$ so that a black hole may eventually form.

Although these estimates are encouraging, the evolution of high-mass
X-ray binaries is complicated and not well understood.
Terman, Taam, \& Hernquist (1995) 
have demonstrated that for a wide range of orbital periods and companion
masses, the spiral-in is rapid and the common envelope can be ejected 
by the interaction of the neutron star with its red supergiant
companion, resulting in a neutron star-helium core binary, 
not a black hole. However, for a sufficiently 
short orbital period of $P<P_{crit}$, where $P_{crit}$ is the critical 
orbital period of the progenitor binary ranging from 0.2 yr to 2 yr 
for companion masses  $12 M_{\odot}$-$24 M_{\odot}$, Terman et al.
(1995) showed that the system would eventually merge to 
form a red supergiant with 
a neutron star core, namely, a massive Thorne-Zytkow Object (TZO)  
(Thorne \& Zytkow 1977). 

Whether AICNS will eventually occur in a TZO is unclear.
The evolution of massive TZOs was addressed by Podsiadlowski, Cannon, 
\& Rees (1995), who found that stellar-mass black holes are likely 
to form as one of the three endpoints of TZOs. A similar conclusion
was reached by Fryer, Benz, \& Herant (1996), who argued,
based on a numerical study of rapid neutron star accretion,
that TZOs may lead to the super-Eddington 
accretion onto neutron stars and the formation of black holes.  
Podsiadlowski et al. (1995) also estimated a TZO birthrate of
$\geq 10^{-4}$ yr$^{-1}$ in the Galaxy, which is not significantly
different from that required for GRBs if we notice the fact that 
a number of TZOs may survive during the spiral-in phase and will not
give rise to AICNS (Bhattacharya \& van den Heuvel 1991).
However, Podsiadlowski et al. (1995) pointed out that 
the rapid accretion model of Chevalier (1993) assumes spherical
symmetry, which is unlikely, and a more realistic calculation 
might rule out AICNS in TZO. 
 
Taking these arguments as a whole, we feel that 
much work will be needed to understand the evolutionary
scenario of high-mass X-ray binaries and to improve 
our naive model of AICNS proposed in the present letter.

The fate of a neutron star as a result of collapse is the formation of 
a stellar black hole. The binary system may still survive 
if the AICNS occurs before it merges with the core of
its companion and if the newly born black hole and the core (probably
a newly born neutron star) remain
gravitationally bound. It is also possible that an isolated 
black hole is left if the explosion expels all the substance from 
the companion. A birthrate of $\sim10^{-5}$ yr$^{-1}$ in the Galaxy for
the AICNS would result in about $10^5$ stellar black holes, 
including the black hole-neutron star binaries. This number is
comparable to the estimate of the population of double neutron star and
black hole-neutron star binaries in the Galaxy by Narayan et al. (1991).

As in the hypernova model (Paczy\'nski 1997), the AICNS in a massive
binary will lead to a ``dirty'' fireball since the explosion is most 
likely to take place inside the gas envelope of the companion star, where
matter ejection following the explosion is non-spherical 
if the AICNS does not occur at the center of the companion.
Consequently, in contrast to the hypernova model, 
here a non-spherical expansion appears to
be a natural result, which may eventually accelerate a small 
fraction of matter to a very large Lorentz factor, creating 
the GRBs.  Since the bulk of the ejecting matter in the companion star
moves slowly,  the afterglow may show up for a relatively long time
after the observation of the GRB. It appears that the two arguments
raised explicitly by Vietri (1997b) can be easily understood in the
framework of the AICNS scenario: (1) A nonuniform external medium
surrounding the explosion is needed to explain the existence 
of a delay between the optical luminosity maximum and the GRB;
and (2) the energy source powering the GRB should be the same as that
of the afterglow.

\section{Conclusions}

In a high-mass X-ray binary system, a significant mass accretion  
onto the neutron star may take place in the later
evolutionary phase of the O/B companion star. If the mass of a
neutron star exceeds the limit 
$M_{max}\approx3M_{\odot}$, the neutron star will inevitably collapse 
into a black hole. Such an AICNS scenario can be regarded as the  
natural energy source for 
GRB, releasing a total binding gravitational energy of $\sim10^{54}$
erg. Because the explosion is non-spherical inside the companion star, 
a moderate amount of beaming is naturally expected and a small
fraction ($\sim0.1\%$) of the energy will be sufficient to 
account for the GRB. In particular, the bulk of the ejecting matter
of the companion star may give rise to the afterglow. 
It is estimated that the AICNS event rate within redshift $z=1$ is
about $2$ per day, in accord with the currently 
observed rate of GRBs. As compared with other popular models for the 
energy sources of GRBs, namely, the merger of double neutron stars
or a neutron star and a black hole binary and the hypernovae, the AICNS
in a massive binary appears to be more reasonable  according to the
stellar evolutionary scenario, and accounts naturally for a number of
observed features of GRBs. It is pointed out that the AICNS model,
though very schematic at the present stage, may deserve to be considered
seriously in the theoretical study of the origin of the GRBs. In particular, 
one needs to understand the detailed mechanism of how a significant
amount of matter is accreted onto the surface of a neutron star.

\acknowledgments

We thank an anonymous referee for valuable comments and suggestions.
This work was supported by the National Science Foundation of China, 
the RGC Earmarked Grant CUHK4189/97P  and a CUHK direct grant 702/1.  





\begin{references}
\reference{}Bhattacharya, D., \& van den Heuvel, E. P. J. 1991,
		Physics Reports, 203, 1
\reference{}Blinnikov, S. I., Novikov, I. D., Perevodchikova, T. V.,
		\& Polnarev, A. G. 1984, Soviet Astron. Lett., 10, 177
\reference{}Canal, R., Isern, J., \& Labay, J. 1990, \araa, 28, 183
\reference{}Cavallo, G., \& Rees, M. J. 1978, \mnras, 183, 359
\reference{}Chevalier, R. A. 1989, \apj, 346, 847
\reference{}Chevalier, R. A. 1993, \apj, 411, L33
\reference{}Costa, E. et al. 1997, \nat, 387, 783
\reference{}Dar, A., Kozlovsky, B.Z., Nussinov, S., \& Ramaty, R. 1992, 
                \apj, 388, 164
\reference{}Delgado, A. J. 1980, \aap, 87, 343
\reference{}Eichler, D., Livio, M., Piran, T., \& Schramm, D. 1989,
		\nat, 340, 126
\reference{}Fryer, C. L., Benz, W., \& Herant, M. 1996, \apj, 460, 801
\reference{}Fukugita, M., \& Turner, E. L. 1991, \mnras, 253, 99
\reference{}Goodman, J., Dar, A., \& Nussinov, S. 1987, \apj, 314, L7
\reference{}Guarnieri, A. et al. 1997, preprint astro-ph/9707164
\reference{}Houck, J. C., \& Chevalier, R. A. 1991, \apj, 376, 234 
\reference{}Kippenhahn, R., \& Weigert, A. 1990, Stellar Structure
		and Evolution (Springer-Verlag:Berlin)
\reference{}Lipunov, V. M. 1992, Astrophysics of Neutron Stars 
		(Springer-Verlag:Berlin)
\reference{}Lipunov, V. M., Postnov, K. A., Prokhorov, M. E., \&
            Panchenko, I. E. 1995, \apj, 454, 593 
\reference{}Meegan, C. A. et al. 1996, \apjs, 106, 65
\reference{}Metzger, M. R. et al. 1997, IAU Circular No. 6655
\reference{}Meurs, E. J. A., \& van den Heuvel, E. P. J. 1989, \aap,
             226, 88
\reference{}Narayan, R., Paczy\'nski, B., \& Piran, T. 1992, \apj, 395,
             L83
\reference{}Narayan, R., Piran, T., \& Shemi A. 1991, \apj, 379, L17
\reference{}Paczy\'nski, B. 1986, \apj, 308, L43
\reference{}Paczy\'nski, B.. 1991, Acta Astron., 41, 257
\reference{}Paczy\'nski, B. 1992, \nat, 355, 521
\reference{}Paczy\'nski, B. 1997, \apj, submitted (astro-ph/9706232)
\reference{}Podsiadlowski, P., Cannon, R. C., \& Rees, M. J. 1995, 
            \mnras, 274, 485
\reference{}Rees, M. J., \& M\'esz\'aros, P. 1992, \mnras, 258, 41P
\reference{}Sahu, K. C., et al. 1997, \nat, 387, 476
\reference{}Taam, R. E., Bodenheimer, P., \& Ostriker, J. P. 1978, 
            \apj, 222, 269 
\reference{}Terman, J. L., Taam, R. E., \& Hernquist, L. 1995, \apj,
            445, 367 
\reference{}Thorne, K. S., \& Zytkow, A. N. 1977, \apj, 212, 832
\reference{}van Paradijs, J. et al. 1997, \nat, 386, 686
\reference{}Verbunt, F. 1993, \araa, 31, 93
\reference{}Vietri, M. 1997a, \apj, 478, L9
\reference{}Vietri, M. 1997b, \apj, submitted (astro-ph/9706060)
\reference{}Waxman, E., 1997, \apj, 485, L5 
\reference{}Woosley, S. E. 1993, \apj, 405, 273
\end{references}
\end{document}